\documentclass[pss]{wiley2sp}
\usepackage{amsmath}
\usepackage{amssymb}
\usepackage{graphicx}
\usepackage{hyperref}
\usepackage{color,xcolor}

\begin{document}

\title{Block antiferromagnetism and possible ferroelectricity in KFe$_2$Se$_2$}

\titlerunning{Block antiferromagnetism ...}

\author{Yang Zhang, Huimin Zhang, Yakui Weng, Lingfang Lin, Xiaoyan Yao, Shuai Dong\textsuperscript{\Ast}}
% Abbreviated list of authors for the page headers
\authorrunning{Y. Zhang et al.}

\mail{e-mail
  \textsf{sdong@seu.edu.cn}, Phone:
  +86-025-52090606-8207, Fax: +86+025-52090600-8201}
\institute{Department of Physics, Southeast University, Nanjing 211189, China}
\received{XXXX, revised XXXX, accepted XXXX} % do not change, will be filled in by the publisher
\published{XXXX} % do not change, will be filled in by the publisher

\keywords{ferrielectric/antiferroelectric; antiferromagnetism; iron-selenide}

\abstract{
\abstcol{Superconductors and multiferroics are two of the hottest branches in condensed matter physics. The connections between those two fields are fundamentally meaningful to unify the physical rules of correlated electrons. Recently, BaFe$_2$Se$_3$, was predicted to be multiferroic [Phys. Rev. Lett. 113, 187204 (2014)] due to its unique one-dimensional block-type antiferromagnetism. Here, another iron-selenide KFe$_2$Se$_2$, a parent state of iron-based superconductor, is predicted to be multiferroic. Its two-dimensional block-type antiferromagnetism can generate a moderate electric dipole for each Fe-Se layer via the Fe-Se-Fe exchange striction. Different stacking configurations of these magnetic blocks give closely proximate energies and thus the ground state of KFe$_2$Se$_2$ may be switchable between antiferroelectric and ferroelectric phases.}}

\titlefigure{abstract}
\titlefigurecaption{Crystal structure of KFe$_2$Se$_2$. (a) Purple: K; green: Se; brown: Fe. Two Fe sheets in a minimum unit cell are indicated as A and B. (b) One Fe-Se layer with magnetism. Brown: spin up; blue: spin down. (c) A side view of Fe-Se bonds. The ionic displacements driven by exchange striction are indicated by arrows.}

\maketitle

\section{Introduction}
Superconductivity and superconductors have always been the most attractive topic in condensed matter physics, while magnetoelectricity and multiferroics have become another emergent hotspot since 2003. Both these two fields are not only physical meaningful to understand the correlated electrons in solids, but also technological important for applications. In spite of the conceptual difference between these two fields, the studies of high-temperature superconductors can teach important information to the multiferroic community. In early years, magnetism was widely believed to contradict superconductivity. But in cuprates, iron-pnictides, and iron-chalcogenides, magnetism plays a crucial role to induce superconductivity \cite{Dagotto:Rmp94,Johnston:Ap,Lumsden:Jpcm,Dagotto:Rmp,Dai:Rmp}. Similar situation also occurs in multiferroics. In the beginning, magnetism was thought to contradict ferroelectricity \cite{Hill:Jpcb}, but in the type-II multiferroics, particular magnetism can induce ferroelectric polarization ($P$) \cite{Cheong:Nm,Dong:Mplb,Dong:Ap}. Thus, magnetism plays as a common ingredient in both high-temperature superconductors and type-II multiferroics, and may link these two branches of correlated electrons.

In fact, there are several cuprates and cupric oxides confirmed to be multiferroics \cite{Kimura:Nm}. And a recent theoretical prediction extended this crossover of superconductor and multiferroics to iron-selenide family  \cite{Dong:PRL14}. In the 123-type BaFe$_2$Se$_3$, the block-type antiferromagnetism can generate an electric dipole moment due to the magnetostriction effect, In contrast, the isostructural BaFe$_2$S$_3$ was found to host superconductivity \cite{Takahashi:Nm}.

Besides the 123-type, there are other types of iron-based superconductors, e.g. the 122-type and its variants (245-type and 234-type) with ordered vacancies \cite{Guo:Prb,Bao:Cpl,Ye:Prl,WangMeng:Prb,Zhao:Prl}. Many experiments argued that in $A_{1-x}$Fe$_{2-y}$Se$_2$ ($A$: alkaline earth) the insulating antiferromagnetic (AFM) and superconducting phases were coexisting, namely the system was phase separated \cite{WangMeng:Prb,Yan:Sr,Yuan:Sr,Li:Np,Zhang:Sr13}. The specific AFM order in the 245-type K$_{0.8}$Fe$_{1.6}$Se$_2$ is also a kind of block-type antiferromagnetism, but with ordered vacancies \cite{Yu-Rong:Prb,Fang:Prb,Lu:Cpb}. The parent phase, i.e. stoichiometric KFe$_2$Se$_2$, is difficult to be synthesized in bulk, but has been observed in thin films \cite{Li:Np,Li:Prl}. A density functional theory (DFT) calculation predicted it to own two-dimensional block-type AFM order \cite{Li:Prb12}, which was verified indirectly in experiment as the observation of block-type structural distortion \cite{Li:Np}. 
Inspired by the 123-type BaFe$_2$Se$_3$ \cite{Dong:PRL14}, the 122-type KFe$_2$Se$_2$ may extend the crossover between superconductors and multiferroics.

Structurally, KFe$_2$Se$_2$ forms the tetragonal crystal structure, whose space group is $I4/mmm$ (No. 139) (see abstract). In each unit cell, there are two Fe layers, each of which is built by edge-sharing FeSe$_4$ tetrahedra. K ions intercalate between Fe-Se layers. In this work, the magnetic order and magnetostriction of KFe$_2$Se$_2$ will be studied using the DFT calculations to explore possible ferroelectricity. Our study indeed finds the dipole moment of each Fe-Se layer, while the global ferroelectric polarization is very sensitive to perturbation.

\section{Method}
The DFT electronic structure calculations are performed using the Vienna {\it ab initio} Simulation Package (VASP) with the projector augmented-wave (PAW) potentials based on the generalized gradient approximation (GGA) \cite{Kresse:Prb,Kresse:Prb96,Blochl:Prb,Perdew:Prl}. To accurately describe the structure of possible ferroelectric distortion, the revised PBE (PBEsol) function is adopted \cite{Perdew:Prl08}. The plane-wave cutoff is 500 eV.

It is well known that DFT usually underestimates band gaps. To overcome this drawback, both the GGA+$U$ method using the Dudarev implementation \cite{Dudarev:Prb} and the hybrid functional calculation based on Heyd-Scuseria-Ernzerhof (HSE) exchange \cite{Heyd:Jcp,Heyd:Jcp04,Heyd:Jcp06} are also adopted.

\begin{figure}[t]%
\vskip -0.55 cm
\includegraphics*[width=\linewidth] {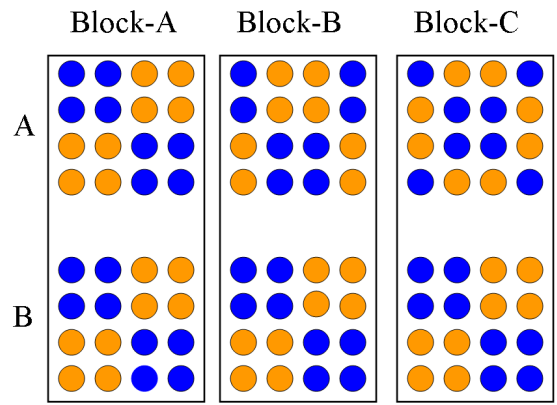}
\caption{Sketch of the block AFM series. A and B denote the two layers shown in abstract(a). Left: Block-A. Middle: Block-B. Right: Block-C. Irons with spin up and spin down are distinguished by colors.}
\label{Fig1}
\end{figure}

To accommodate various magnetic orders, a $2\times$2$\times1$ supercell is used. The $k$-point mesh is $4\times$4$\times2$. Both the lattice constants and inner atomic positions are fully relaxed until the force on each atom is below $0.01$ eV/{\AA}. Several magnetic orders have been considered, including the conventional ferromagnetic (FM), stripy-AFM (C-AFM), checkerboard-AFM (G-AFM), as well as block-AFM and its variants (Fig.~\ref{Fig1}).

\section{Results}
\subsection{Symmetry analysis}
The exchange striction effect is active for the block-AFM order. According to the previous DFT study \cite{Li:Prb12}, there is prominent tetramerization of irons driven by its block-AFM ordering, which has been confirmed by scanning tunnel microscope (STM) \cite{Li:Np}. In details, the nearest-neighbor (NN) distance between Fe($\uparrow$) and Fe($\uparrow$) [or Fe($\downarrow$)-Fe($\downarrow$)] is shorten comparing with the one between Fe($\uparrow$)-Fe($\downarrow$), similar to the situation in BaFe$_2$Se$_3$ \cite{Dong:PRL14,Caron:Prb}. However, only this kind of exchange striction of Fe-Fe pairs can not break the space-inversion symmetry. The staggered (upper-lower-upper-lower) configuration of Se is crucial to couple with the block-AFM order. Thus, the exchange striction of Fe-Se-Fe bonds is responsible for the ferroelectric dipoles.

Microscopically, as shown in abstract(b-c), Se(4) and Se(5) are located in the opposite sides of iron layer. Originally, the distance of Se(4) and Se(5) to the iron layer should be identical. Due to the tetramerization, the shrunk Fe($\uparrow$) blocks will push Se(4) ion upward, while Se(5) will be lifted up due to the elongation of Fe($\uparrow$)-Fe($\downarrow$) bond. Similar movements occur for other Se ions. As a result, the movements of Se ions break the inversion symmetry and generate a local dipole moment pointing perpendicular to the iron plane, i.e. along the $c$ axis.

Although above analysis suggests that each Fe-Se layer should be multiferroic, this layered system can be either a ferroelectric one with a finite macroscopic ferroelectric $P$ or an antiferroelectric one with canceled $P$, depending on the stacking of magnetic blocks along the $c$-axis.

Formal group theory can also confirm the ferroelectricity of block-AFM order. %The space group is No. 139 ($I4/mmm$, Tetragonal) for the pure lattice structure. Using this highly symmetric structure and imposing the magnetism,
The magnetic+lattice space group changes from No. 139 ($I4/mmm$) to: (a) No. 51 ($Pmma$ for Block-A); (b) No. 36 ($Cmc2_1$ for Block-B); (c) No. 123 ($P4/mmm$ for Block-C). The corresponding point groups of these three space groups are $mmm$, $mm2$, and $4/mmm$, among which only the $mm2$ is a polar point group. The point group for relaxed structure of Block-B deceases to $m$, which is still a polar point group and allows the ferroelectric $P$.

\subsection{DFT results}
Although a previous DFT study already studied the structural and magnetic properties of KFe$_2$Se$_2$  \cite{Li:Prb12,Cao:Cpl}, it is necessary to re-check these results. In particular, the Block-type magnetism was not considered in Ref. [41]. Although in Ref. [26] the Block state was considered, the calculated lattice constants are largely divergent from the experimental values.
For example, the LDA+$U$ relaxed lattice constants ($a=3.98$ {\AA}, $c=14.39$ {\AA}) \cite{Li:Prb12} for the ground magnetic state is much larger than the experimental one $a=3.89$ {\AA}, $c=14.10$ {\AA}) \cite{Li:Np}. In general, for transition metal compounds, it is well known that the magnetism and ferroelectric distortions may seriously depend on the structure. In the present study, the choice of PBEsol function can give a more accurate description of structure, which is nontrivially important for reliability of predictions, especially regarding the ferroelectricity.

\begin{figure}[t]%
\includegraphics*[width=\linewidth]{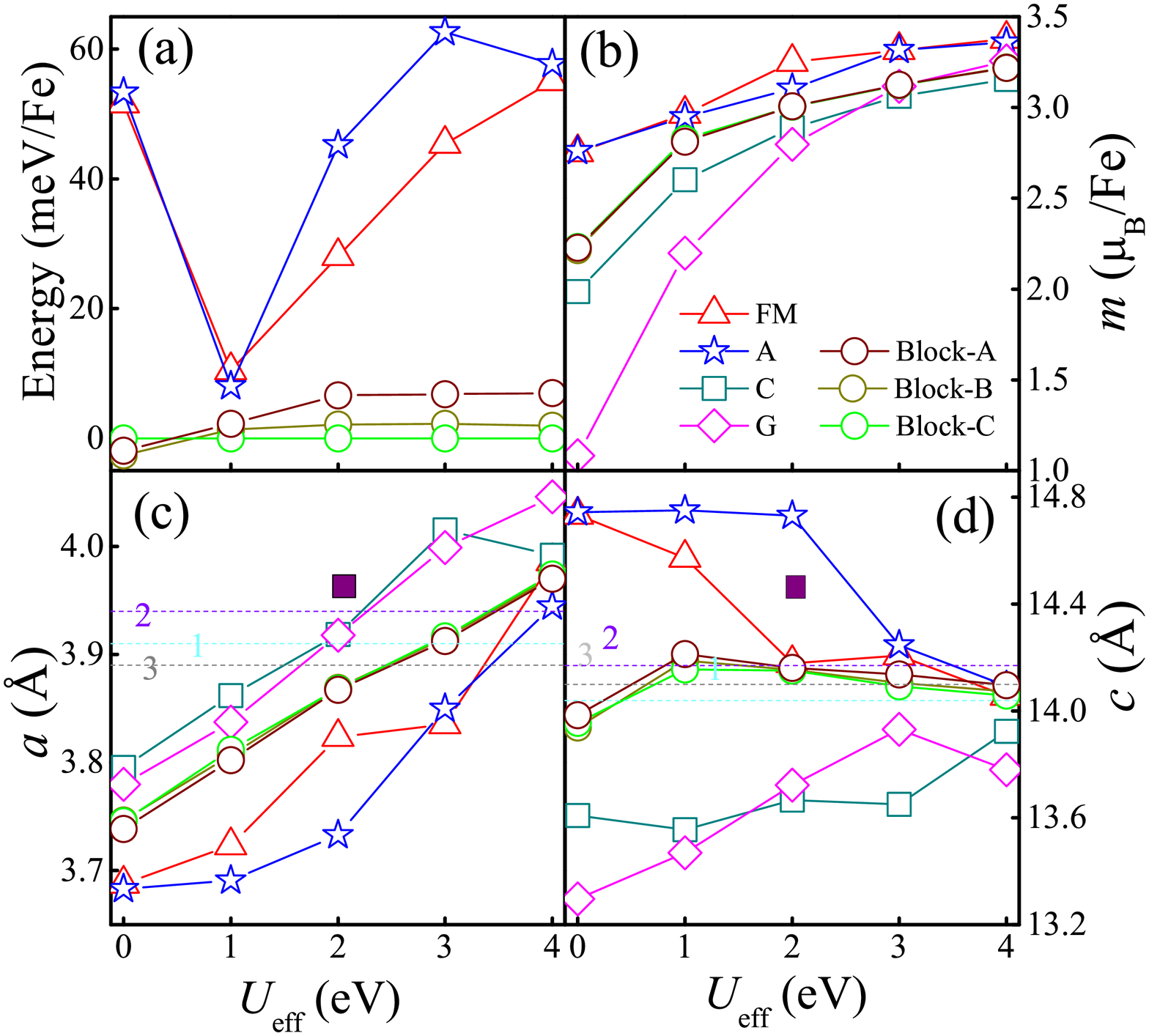}
\caption{DFT results for various magnetic orders as a function of $U_{\rm eff}$. (a) Energies. The Block-\textbf{C} state is taken as the reference. (b) Local magnetic moments of Fe. (c-d) The optimized lattice constants along the $a$-axis and $c$-axis. The broken lines (1 to 3) correspond to the experimental values \cite{Guo:Prb,Bao:Cpl,Li:Np}. Noting some experimental values were measured at room temperature, and vacancies existed in their samples. Solid square: previous DFT value \cite{Li:Prb12}.}
\label{Fig2}
\end{figure}

First, both the lattice constants and atomic positions were optimized with various magnetic structures by varying the effective Hubbard interaction $U_{\rm eff}$. Based on relaxed structures, the energies of various magnetic orders are compared, as shown in Fig.~\ref{Fig2}(a). The block AFM series always own lower energies than others. This conclusion is robust, independent of the value of $U_{\rm eff}$. Within the block AFM series, the ground state changes from the Block-B to Block-C with increasing $U_{\rm eff}$. The energy differences within the block AFM series are very tiny, about $2-6$ meV/Fe. It is reasonable considering the quasi-two-dimensional Fe layers. Thus, the almost degenerated block AFM series may provide switchable functions via external stimulates.An external electric field should be applied along the c axis, such as using Piezoresponse Force Microscopy (PFM) trips to apply a local electric field. If a large enough field is applied along the c axis, the antiferroelectric (Block-C) to FE (Block-B) phase transition will occur, producing a 180¡ã flipping and enhancement of $P$.

Second, the local magnetic moment of Fe is displayed in Fig.~\ref{Fig2}(b). The block AFM series own almost identical magnetic moment, which increases from $2.2$ $\mu_B$/Fe to $3.2$ $\mu_B$/Fe with $U_{\rm eff}$. Such a large moment may imply a non-superconductive fact for this $\sqrt2\times\sqrt2$-type distorted KFe$_2$Se$_2$, in agreement with the STM measurement \cite{Dagotto:Rmp,Li:Np}. Of course, it should be noted that the calculated magnetic moment using DFT + U is always systematic overestimated, slightly larger than the LDA + DMFT results (e.g. $2$ $\mu_B$ /Fe) \cite{Yin:Nm}.

Third, the relaxed lattice constants are summarized in Fig.~\ref{Fig2}(c-d). The experimental values and previous DFT value are also shown for comparison. It is clear that our calculation with the PBEsol function can give much better description of KFe$_2$Se$_2$. For example, at $U_{\rm eff}=2$ eV, our optimized lattice constants ($a=3.87$ {\AA}, $c=14.15$ {\AA}), are very close to the experimental one. The accurate description of structure is a good start point for further calculation on ferroelectricity.

\begin{figure}[t]%
\includegraphics*[width=\linewidth]{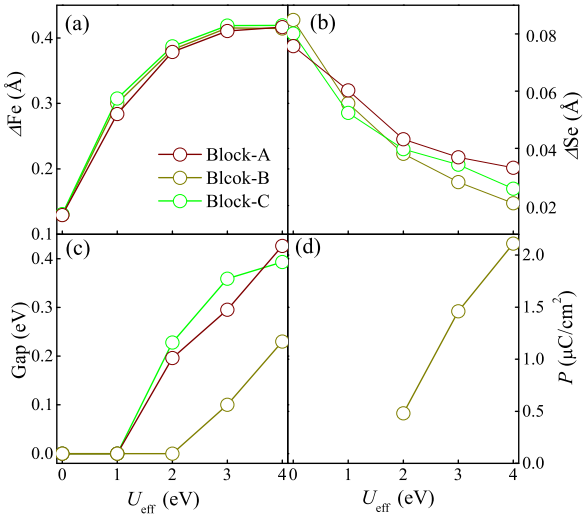}
\caption{DFT results as a function of $U_{\rm eff}$. (a) $\Delta$Fe: the difference between elongated and shrunken Fe¨CFe distances. (b) $\Delta$Se: the Se's centroid height to the Fe sheet for an individual FeSe layer. (c) Band gaps. Only block states are shown while other states are metallic. (d) The ferroelectric polarization of Block-B state.}
\label{Fig3}
\end{figure}

Fourth, the magnetostrictive effect in the block AFM series can be illustrated in Fig.~\ref{Fig3}(a-b). The bond lengths are indeed different between Fe($\uparrow$)-Fe($\uparrow$) [or Fe($\downarrow$)-Fe($\downarrow$)] and Fe($\uparrow$)-Fe($\downarrow$), e.g. $2.542$ {\AA} and $2.929$ {\AA} respectively when $U_{\rm eff}=2$ eV. As a side effect of this exchange striction, the heights of Se (to the iron plane) becomes different: $1.53$ {\AA} for Se(4) and $1.58$ {\AA} for Se(5), respectively. Thus, the individual Fe-Se layer should be polar. Surprisingly, $\Delta_{\rm Se}$ and $\Delta_{\rm Fe}$ own different behaviors upon increasing $U_{\rm eff}$, in opposite to the intuitional expectation, which is due to the partial covalent Fe-Se bonds (to be discussed later).

Fifth, the band gaps of various magnetic states (Fig.~\ref{Fig3}(c)) show that only those block AFM series are insulating when $U_{\rm eff}\geq2$ eV. To avoid the indeterminacy from the choice of $U_{\rm eff}$, the result of HSE calculation is also shown, which indicates an insulating behavior with a gap of \textbf{0.65} eV.

Finally, to obtain a net $P$, it is essential to have parallel alignment of dipole moments of nearest-neighbor layers. Our DFT calculations indeed find a finite $P$ for the Block-B, while other two block-AFM's are antiferroelectric. For the Block-B, the net $P$ calculated using the standard Berry phase method \cite{King-Smith:Prb,Resta:Rmp} is along the $c$-axis as expected and its magnitude increases from $0.48$ $\mu$C/cm$^2$ ($U_{\rm eff}=2$ eV) to $2.1$ $\mu$C/cm$^2$ ($U_{\rm eff}=4$ eV). This magnitude is considerable prominent among type-II multiferroics \cite{Cheong:Nm,Dong:Mplb,Dong:Ap}.

In summary of this subsection, our DFT calculations confirm the block-type antiferromagnetism as the ground state and the magnetism-induced ferroelectricity (or antiferroelectricity) of KFe$_2$Se$_2$. Our prediction is not limited to KFe$_2$Se$_2$ but can be naturally extended to other $A$Fe$_2$Se$_2$ ($A$=Cs, Rb, Tl), considering the same crystal structure and similar electron structure.

\section{Discussion}
According to the calculated density of states (not shown here), the bands near the Fermi levels are mostly contributed by Fe's $3d$ orbitals, which are highly hybridized with Se's $4p$ orbitals. The hybridization implies the partial covalency of Fe-Se bonds, as previously argued in BaFe$_2$Se$_3$ \cite{Dong:PRL14}. %For comparison, the HSE exchange are performed in Fig. 4(c).The different prediction of parameters have been calculated. The ideal insulating phase  have been obtained in different prediction. For simplicity, the high accurate results are shown in Fig 4(c)which also shows a highly hybridized between Fe's $3d$ orbitals and Se's $4p$ orbitals.

In fact, the intuitional point charge model can also give a rough estimation of polarization. Taking the relaxed structure at $U_{\rm eff}=2$ eV and assuming K$^{1+}$, Fe$^{1.5+}$ and Se$^{-2}$, the point charge model gives $0.86$ $\mu$C/cm$^2$, which is larger than the Berry phase value. It is reasonable since the values of Born effective charge are usually lower than their nominal valences, especially considering the weak electronegativity of Se. Ref.~\cite{Dong:PRL14} once argued that the effective valence of Fe in KFe$_2$Se$_2$ is possible $+1.25$ or $+0.75$ considering its tetramerized block magnetism. The $+0.75$ valence of Fe (and thus $-1.25$ for Se) indeed gives better consistent between the point charge model ($0.54$ $\mu$C/cm$^2$ for the $U_{\rm eff}$=2 eV case) and the Berry phase one. Furthermore, the high spin state of Fe$^{0.75+}$ gives a magnetic moment $2.75$ $\mu_B$/Fe, which is also very close to the calculated local moment (shown in Fig.~\ref{Fig2}(b)). %These hypotheses can be also quasi-quantitatively confirmed by the Bader charge analysis,\cite{Tang:Jpcm,Bader:book,Henkelman:Cms} as shown in Fig. 4(d).

The point charge model can be used to analyze the layer dipole moment of antiferroelectric Block-A and Block-C, which can not be directly estimated by the Berry phase method. This dipole moment is about $0.43\times V_{\rm Se}$ $\mu$C/cm$^2$ (at $U_{\rm eff}=2$ eV), pointing along the $c$-axis. The dipole moments between any nearest-neighbor layers are aligned antiparallelly, rendering the antiferroelectric fact.

Till now, the stoichiometric KFe$_2$Se$_2$ is difficult to obtain in experiment, especially in the bulk form. The intrinsic reason is probably due to low effective valence of Fe, e.g. $+0.75$ mentioned before, which is unstable. Thus a natural idea is to replace Se by S, which can enhance the electronegativity and thus the Fe¡¯s effective valence. Although to our best knowledge there is no report on KFe$_2$S$_2$ till now, a DFT calculation can be done for comparison with KFe$_2$Se$_2$. Our calculation found a quite different magnetic ground state for KFe$_2$S$_2$. The rare bicollinear-AFM gives the lowest energy, when comparing with other magnetic candidates including FM, A-AFM, C-AFM, block-AFMs, and G-AFM. In the family of iron-based superconductors, this bicollinear one only appears in FeTe \cite{Subedi:Prb}. The nominal valence of Fe in FeTe is $+2$, which indeed higher than the one in KFe$_2$Se$_2$. This tendency agrees with the physics in KFe$_2$S$_2$. Following this prediction and analysis, it is also possible to obtain the bicollinear-AFM by reducing the amount of K in KFe$_2$S$_2$ to some extend.

\section{Conclusion}
The structural, magnetic, ferroelectricity, and electronic structure of KFe$_2$Se$_2$ have been systematically investigated using the first-principles calculations. The block-type antiferromagnetic series were unambiguously confirmed to be the ground state, although the stacking configuration along the $c$-axis is quite subtle. The magnetism driven dipole moment was verified for each Fe layer, while the total polarization depends on the stacking configuration along the $c$-axis. Antiferroelectric-ferroelectric transition is possible via proper stimulates. Our study could be naturally extended to cover other $A$Fe$_2$Se$_2$ ($A$=Cs, Rb, Tl) considering the same crystal structure and similar electron structure, while the isostructural $A$Fe$_2$S$_2$ was predicted to show exotic bicollinear magnetism.

\begin{acknowledgement}
We thank Y. H. Li, J. J. Zhang, and B. Gao for helpful discussions. This work was supported National Natural Science Foundation of China (Grant No. 51322206) and Natural Science Foundation of Jiangsu Province of China (Grant No. BK20141329).
\end{acknowledgement}

\end{document}